\documentclass[aps,prr,twocolumn,superscriptaddress,nofootinbib]{revtex4}

\usepackage{graphicx}
\usepackage{dcolumn}
\usepackage{bm}
\usepackage{amssymb}
\usepackage{amsfonts}
\usepackage{latexsym}
\usepackage{enumerate}
\usepackage[dvipdfmx]{color} 
\usepackage{amsmath}
\usepackage[dvipdfmx]{epsfig}
\usepackage{times}
\usepackage{algorithm}
\usepackage{algorithmic}
\usepackage{cases}

\newcommand{\red}{\color{black}}
\newtheorem{theorem}{Theorem}
\newtheorem{lemma}{Lemma}
\newtheorem{definition}{Definition}

\begin{document}
\widetext
\title{\red Hardness of efficiently generating ground states in postselected quantum computation}
\author{Yuki Takeuchi}
\email{yuki.takeuchi.yt@hco.ntt.co.jp}
\affiliation{NTT Communication Science Laboratories, NTT Corporation, 3-1 Morinosato Wakamiya, Atsugi, Kanagawa 243-0198, Japan}
\author{Yasuhiro Takahashi}
\affiliation{NTT Communication Science Laboratories, NTT Corporation, 3-1 Morinosato Wakamiya, Atsugi, Kanagawa 243-0198, Japan}
\author{Seiichiro Tani}
\affiliation{NTT Communication Science Laboratories, NTT Corporation, 3-1 Morinosato Wakamiya, Atsugi, Kanagawa 243-0198, Japan}

\begin{abstract}
Generating ground states of any local Hamiltonians seems to be impossible in quantum polynomial time.
In this {\red paper}, we give evidence for the impossibility by applying an argument used in the quantum-computational-supremacy approach.
More precisely, we show that if ground states of any $3$-local Hamiltonians can be approximately generated in quantum polynomial time with postselection, then ${\sf PP}={\sf PSPACE}$.
Our result is superior to the existing findings in the sense that we reduce the impossibility to an unlikely relation between classical complexity classes.
We also discuss what makes efficiently generating the ground states hard for postselected quantum computation.
\end{abstract}
\maketitle

{\red\section{Introduction}}
Quantum computing is expected to outperform classical computing.
Indeed, quantum advantages have already been shown in terms of query complexity~\cite{S97} and communication complexity~\cite{R99}.
Regarding time complexity, it is also believed that universal quantum computing has advantages over classical counterparts.
For example, although an efficient quantum algorithm, i.e., Shor's algorithm, exists for integer factorization~\cite{Sh97}, there is no known classical algorithm that can do so efficiently.
However, an unconditional proof that there is no such classical algorithm seems to be hard because an unconditional separation between ${\sf BQP}$ and ${\sf BPP}$ implies ${\sf P}\neq{\sf PSPACE}$.
Whether ${\sf P}\neq{\sf PSPACE}$ is a long-standing problem in the field of computer science.

To give evidence of quantum advantage in terms of computational time, a sampling approach has been actively studied.
This approach is to show that if the output probability distributions from a family of (non-universal) quantum circuits can be efficiently simulated in classical polynomial time, then the polynomial hierarchy (${\sf PH}$) collapses to its second or third level.
Since it is widely believed that ${\sf PH}$ does not collapse, this approach shows one kind of quantum advantage (under a plausible complexity-theoretic assumption).
This type of quantum advantage is called quantum computational supremacy~\cite{HM17}.
The quantum-computational-supremacy approach is remarkable because it reduces the impossibility of an efficient classical simulation of quantum computing to unlikely relations between classical complexity classes {\red(under conjectures such as the average-case hardness conjecture)}.
Since classical complexity classes have been studied for a longer time than quantum complexity classes, unlikely relations between classical complexity classes would be more dramatic than those involving quantum complexity classes.
As subuniversal quantum computing models showing quantum computational supremacy, several models have been proposed, such as boson sampling~\cite{AA11,LLRROR14,HKSBSJ17}, instantaneous quantum polynomial time (IQP)~\cite{BJS10,BMS16} and its variants~\cite{TTYT15,TT16,GWD17,MSM17}, deterministic quantum computation with one quantum bit (DQC1)~\cite{MFF14,FKMNTT18}, Hadamard-classical circuit with one qubit (HC1Q)~\cite{MTN18}, and quantum random circuit sampling~\cite{BISBDJBMN18,HBSE18,BFNV19,MTT19}.
A proof-of-principle demonstration of quantum computational supremacy has recently been achieved using quantum random circuit sampling with 53 qubits~\cite{google}.
Regarding other models, small-scale experiments have been performed toward the goal of demonstrating quantum computational supremacy~\cite{LBAW08,SMHKJBDTLKGSSW13,TDHNSW13,CORBGSVMMS13,BSVFVLMBGCROS15,WQDCCYHJYWSRHLP19}.

On the other hand, the limitations of universal quantum computing are also actively studied (e.g., see Refs.~\cite{B09,M17,GH20}).
Understanding these limitations is important to clarify how to make good use of universal quantum computers.
For example, it is believed to be impossible {\red in the worst case} to generate ground states of {\red a given local Hamiltonian} in quantum polynomial time, while their heuristic generation has been studied using quantum annealing~\cite{KN98}, variational quantum eigensolvers (VQE)~\cite{PMSYZLGO14}, and quantum approximate optimization algorithms (QAOA)~\cite{FGG14}.
Since deciding whether the ground-state energy of a given $2$-local Hamiltonian is low or high with polynomial precision is a ${\sf QMA}$-complete problem~\cite{KKR06}, if efficient generation of the ground states is possible, then ${\sf BQP}={\sf QMA}$ that seems to be unlikely.
As well as the gap between quantum and classical computing in terms of time complexity, it is hard to unconditionally show the impossibility of efficiently generating the ground states.

In this {\red paper}, we utilize a technique from the quantum-computational-supremacy approach to give new evidence for this impossibility.
More precisely, in Theorem~\ref{additive}, we show that if the ground states of any {\red given} $3$-local Hamiltonians can be approximately generated in quantum polynomial time with postselection, then ${\sf PP}={\sf PSPACE}$. Similar to the quantum-computational-supremacy approach, this consequence leads to the collapse of a hierarchy, i.e., the counting hierarchy (${\sf CH}$) collapses to its first level (${\sf CH}={\sf PP}$).
In Theorem~\ref{inferiority}, we consider a different notion of approximation and show that if the probability distributions obtained from the ground states can be approximately generated in quantum polynomial time with postselection, then ${\sf PP}={\sf PSPACE}$.
Theorem~\ref{inferiority} studies the hardness of approximately generating the ground states from a different perspective, because it is closely related to the hardness of approximately generating the probability distributions.
Furthermore, by using a similar argument, we show that if the ground states of any $3$-local Hamiltonians can be {\red specified} by {\red using polynomial numbers} of bits, then ${\sf NP}^{\sf PP}={\sf PSPACE}$. This leads to the second-level collapse of ${\sf CH}$, i.e., ${\sf CH}={\sf PP}^{\sf PP}$.
This result seems to give additional evidence to support the conclusion that ${\sf QMA}$ is strictly larger than ${\sf QCMA}$.
{\red Here, we say that a ground state is specified by using a polynomial number of bits if there exists a polynomial-size quantum circuit that outputs the ground state given the success of the postselection.
Since ordinary universal gate sets contain only a constant number of elementary gates, we can specify any polynomial-size quantum circuit by using a polynomial number of bits.}
Our results are different from the existing ones on the impossibility of efficient ground-state generation in a sense that we reduce the impossibility to unlikely relations between classical complexity classes as in the quantum-computational-supremacy approach.

{\red This paper is organized as follows.
In Sec.~\ref{II}, we give some preliminaries on complexity classes and show an important lemma (Lemma~\ref{remark}) that is used to obtain our theorems.
In Sec.~\ref{III}, as the first main result, we show that it is hard for postselected quantum computers to approximately generate ground states of a given $3$-local Hamiltonian in the worst case (Theorem~\ref{additive}).
In Sec.~\ref{IV}, as the second main result, we show that it is also hard for postselected quantum computers to approximately generate probability distributions obtained from the ground states, under a different notion of approximation (Theorem~\ref{inferiority}).
Section~\ref{V} is devoted to conclusion and discussion.}

\medskip
{\red\section{Preliminaries}
\label{II}}
Before we explain our results, we will briefly review preliminaries required to understand our argument.
We use several complexity classes that are sets of decision problems.
Here, decision problems are mathematical problems that can be answered by YES or NO.
We mainly use complexity classes ${\sf CH}$, ${\sf postBQP}$, ${\sf postQCMA}$, and ${\sf postQMA}$, where the latter three are postselected versions of ${\sf BQP}$, ${\sf QCMA}$, and ${\sf QMA}$, respectively.
We assume that readers know the major complexity classes, such as ${\sf P}$, ${\sf PP}$, ${\sf PSPACE}$, and ${\sf PH}$ (for their definitions, see Ref.~\cite{AB09}). 

The class ${\sf CH}$ is the union of classes ${\sf C}_k{\sf P}$ over all non-negative integers $k$, i.e., ${\sf CH}=\cup_{k\ge0}{\sf C}_k{\sf P}$, where ${\sf C}_0{\sf P}={\sf P}$ and ${\sf C}_{k+1}{\sf P}={\sf PP}^{{\sf C}_k{\sf P}}$ for all $k\ge 0$.
We say that ${\sf CH}$ collapses to its $k$-th level when ${\sf CH}={\sf C}_k{\sf P}$.
The first-level collapse of {\sf CH} {\red is thought to} be especially unlikely.
This is because, from Toda's theorem~\cite{T91}, ${\sf PH}\subseteq{\sf P}^{\sf PP}\subseteq{\sf CH}$.
Therefore, if ${\sf CH}={\sf PP}$, then ${\sf PH}\subseteq{\sf PP}$.
{\red Although it is unknown whether this inclusion does not hold, it is used as an unlikely consequence in several papers such as Ref.~\cite{V03}.
At least, we can say that it is difficult to show that ${\sf PH}\subseteq{\sf PP}$ holds.
This is because there exists an oracle relative to which ${\sf PH}$ (more precisely, ${\sf P}^{\sf NP}$) is not contained in ${\sf PP}$~\cite{B94}.}

The complexity class ${\sf postQMA}$ is defined as follows~\cite{MN17,UHB17}: a language $L$ is in ${\sf postQMA}$ if and only if there exist a constant $0<\delta<1/2$, polynomials $n$, $m$, and $k$, and a uniform family $\{U_x\}_x$ of polynomial-size quantum circuits, where $x$ is an instance, and $U_x$ takes an $n$-qubit state $\rho$ and ancillary qubits $|0^m\rangle$ as inputs, such that (i) ${\rm Pr}[p=1\ |\ \rho]\ge2^{-k}$, where $p$ is a single-qubit postselection register, for any $\rho$, (ii) if $x\in L$, then there exists a witness $\rho_x$ such that ${\rm Pr}[o=1\ |\ p=1,\rho_x]\ge 1/2+\delta$ with a single-qubit output register $o$, and (iii) if $x\notin L$, then for any $\rho$, ${\rm Pr}[o=1\ |\ p=1,\rho]\le 1/2-\delta$.
In this definition, ``polynomials'' mean the ones in the length $|x|$ of the instance $x$.
Note that ${\sf postQMA}$ is denoted by ${\sf QMA}_{\sf postBQP}$ in Ref.~\cite{MN17}.

\begin{figure}[t]
\includegraphics[width=9cm, clip]{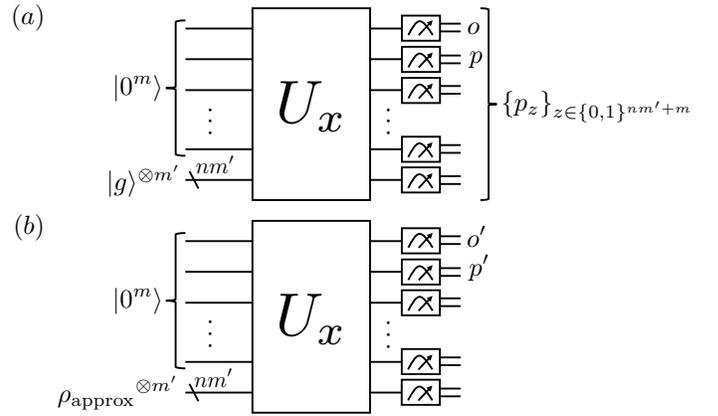}
\caption{(a) A quantum circuit $U_x$ with an input state $|0^m\rangle|g\rangle^{\otimes m'}$ to decide whether $x\in L$ or $x\notin L$, where $L$ is in ${\sf postQMA}$. Let $o$ and $p$ be output and postselection registers, respectively. If $o=p=1$, we conclude that $x\in L$. On the other hand, if $p=1$ and $o=0$, then $x\notin L$. The output probability distribution of $nm'+m$ qubits is denoted by $\{p_z\}_{z\in\{0,1\}^{nm'+m}}$. Each meter symbol represents a $Z$-basis measurement.
(b) The same quantum circuit as in (a) except that $|g\rangle$ is replaced with an approximate state $\rho_{\rm approx}$. The output and postselection registers are denoted by $o'$ and $p'$, respectively.}
\label{multiplicativefig}
\end{figure}

The following is an important lemma:
\begin{lemma}
\label{remark}
Any decision problem in ${\sf postQMA}$ can be efficiently solved using postselected polynomial-size quantum circuits if a polynomial number of copies of a ground state (i.e., a minimum-eigenvalue state) $|g\rangle$ of an appropriate $3$-local Hamiltonian is given [see Fig.~\ref{multiplicativefig} (a)].
Note that a $3$-local Hamiltonian $H=\sum_{i=1}^tH^{(i)}$ with a polynomial $t$ is the sum of polynomially many Hermitian operators $\{H^{(i)}\}_{i=1}^t$, each of which acts on at most three (possibly geometrically nonlocal) qubits.
The operator norm $||H^{(i)}||$ is upper-bounded by one for any $1\le i\le t$.
\end{lemma}
This lemma can be obtained by combining results in Refs.~\cite{MN17,FL18}.
The proof is given in the {\red Appendix}.

By removing $\rho_x$ and $\rho$ from the definition of ${\sf postQMA}$, the complexity class ${\sf postBQP}$ is defined.
Since ${\sf PP}={\sf postBQP}$~\cite{A05}, readers can replace ${\sf PP}$ with ${\sf postBQP}$ if they are not familiar with the definition of ${\sf PP}$.
Furthermore, the class ${\sf postQCMA}$ is defined by replacing each quantum state $\rho_x$ and $\rho$ with a polynomial number of classical bits.
Note that ${\sf postQCMA}$ is denoted by ${\sf QCMA}_{\sf postBQP}$ in Ref.~\cite{MN17}.

\medskip
{\red\section{Hardness of approximately generating ground states}
\label{III}}
We show that efficiently generating approximate ground states of {\red a given} $3$-local {\red Hamiltonian} is hard for postselected quantum computation {\red in the worst case}.
Formally, our first main result is as follows:
\begin{theorem}
\label{additive}
Suppose that it is possible to, for any $n$-qubit $3$-local Hamiltonian $H$ and polynomial $s$, construct a polynomial-size quantum circuit $W$ in classical polynomial time, such that $W$ generates an $n$-qubit state $\rho_{\rm approx}$ given the success of the postselection, satisfying $\langle g|\rho_{\rm approx}|g\rangle\ge1-2^{-s}$ for a ground state $|g\rangle$ of $H$, and the postselection succeeds with probability at least the inverse of an exponential.
Then, ${\sf PP}={\sf PSPACE}$.
\end{theorem}
{\it Proof.}
Our goal is to show that if the quantum circuit $W$ exists, then ${\sf postQMA}\subseteq{\sf postBQP}$.
From ${\sf PP}\subseteq{\sf PSPACE}$, ${\sf postQMA}={\sf PSPACE}$~\cite{MN17}, and ${\sf postBQP}={\sf PP}$~\cite{A05}, this immediately means ${\sf PP}={\sf PSPACE}$.

First, we consider the language $L$ that is in ${\sf postQMA}$.
From Lemma~\ref{remark}, for any instance $x$, there exist polynomials $m$ and $m'$ such that a polynomial-size quantum circuit $U_x$ with input $|0^m\rangle|g\rangle^{\otimes m'}$ efficiently decides whether $x\in L$ or $x\notin L$ under postselection of $p=1$ [see Fig.~\ref{multiplicativefig} (a)].
Here, $|g\rangle$ is a ground state of an $n$-qubit $3$-local Hamiltonian $H_x$ that depends on the instance $x$, $n$ is a polynomial in $|x|$, and $p$ is the postselection register of $U_x$.
From the definition of ${\sf postQMA}$, the postselection succeds with probability ${\rm Pr}[p=1]\ge 2^{-k}$ for a polynomial $k$.

\begin{figure}[t]
\includegraphics[width=7cm, clip]{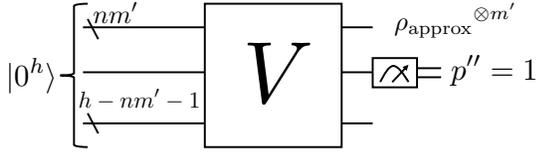}
\caption{A polynomial-size quantum circuit $V$ that prepares tensor products ${\rho_{\rm approx}}^{\otimes m'}$ of an $n$-qubit approximate ground state from $|0^h\rangle$ with polynomials $m'$ and $h(\ge nm'+1)$ when the postselection register $p''=1$. Note that the probability of obtaining $p''=1$ is at least the inverse of an exponential.}
\label{postgsfig}
\end{figure}

Next, we show that the quantum circuit in Fig.~\ref{multiplicativefig} (a) can be simulated using the quantum circuit $W$.
A classical description of $H_x$ can be obtained in polynomial time from the instance $x$.
From the assumption with the Hamiltonian $H_x$ and the polynomials $n$, $m'$, and $k$ described above, we can construct the quantum circuit $W$ such that it prepares the approximate ground state $\rho_{\rm approx}$ whose fidelity $F$ with $|g\rangle$ is $(1-\Theta(2^{-4k}))^{1/m'}$ given the success of the postselection.
By repeated execution of $W$, we can efficiently prepare ${\rho_{\rm approx}}^{\otimes m'}$ given the success of the postselection.
In other words, from the quantum circuit $W$, we can construct a polynomial-size quantum circuit $V$ that generates tensor products ${\rho_{\rm approx}}^{\otimes m'}$ of the approximate ground state in the case of $p''=1$, where $p''$ is the postselection register of $V$ (see Fig.~\ref{postgsfig}).
The fidelity between $|g\rangle^{\otimes m'}$ and ${\rho_{\rm approx}}^{\otimes m'}$ is $F^{m'}=1-\Theta(2^{-4k})$.
When we denote by $r$ the success probability of postselection of $W$, that of $V$ is ${\rm Pr}[p''=1]=r^{m'}$, which is at least the inverse of an exponential.

By combining the quantum circuit $V$ in Fig.~\ref{postgsfig} and $U_x$ in Fig.~\ref{multiplicativefig} (a), we can construct a new quantum circuit $U'_x$, as shown in Fig.~\ref{postuxfig}.
Note that since $U_x$ is in a uniform family of polynomial-size quantum circuits as per the definition of ${\sf postQMA}$, it can be efficiently constructed from the instance $x$.
The postselection register $\tilde{p}'$ of $U'_x$ is equal to $1$ if and only if the postselection registers of $V$ and $U_x$ are both $1$.
In other words, when $\tilde{p}'=1$, the quantum circuit $V$ outputs the correct state ${\rho_{\rm approx}}^{\otimes m'}$, and the quantum circuit $U_x$ is successfully postselected.
Therefore, ${\rm Pr}[o'=1\ |\ p'=1]={\rm Pr}[\tilde{o}'=1\ |\ \tilde{p}'=1]$, where $\tilde{o}'$ is the output register of $U'_x$, and $o'$ and $p'$ are output and postselection registers in Fig.~\ref{multiplicativefig} (b), respectively.
The only difference between Figs.~\ref{multiplicativefig} (a) and \ref{multiplicativefig} (b) is that the input ground states are exact or approximate ones.
Hereafter, we will consider ${\rm Pr}[o'=1\ |\ p'=1]$ instead of ${\rm Pr}[\tilde{o}'=1\ |\ \tilde{p}'=1]$. 

From a property of fidelity (see Theorem 9.1 and Eq.~(9.101) in Ref.~\cite{NC00}), both $|{\rm Pr}[o=p=1]-{\rm Pr}[o'=p'=1]|$ and $|{\rm Pr}[p=1]-{\rm Pr}[p'=1]|$ are upper-bounded by $2\sqrt{1-F^{m'}}$.
{\red Therefore,
\begin{eqnarray*}
{\rm Pr}[o'=1\ |\ p'=1]&=&\cfrac{{\rm Pr}[o'=p'=1]}{{\rm Pr}[p'=1]}\\
&\ge&\cfrac{{\rm Pr}[o=p=1]-2\sqrt{1-F^{m'}}}{{\rm Pr}[p=1]+2\sqrt{1-F^{m'}}}.
\end{eqnarray*}
When $x\in L$, the inequality ${\rm Pr}[o=1\ |\ p=1]\ge1/2+\delta$ holds.
Therefore,
\begin{eqnarray}
\nonumber
{\rm Pr}[o'=1\ |\ p'=1]&\ge&\cfrac{(1/2+\delta){\rm Pr}[p=1]-2\sqrt{1-F^{m'}}}{{\rm Pr}[p=1]+2\sqrt{1-F^{m'}}}\\
\nonumber
&=&\cfrac{1}{2}+\delta-\cfrac{(3+2\delta)\sqrt{1-F^{m'}}}{{\rm Pr}[p=1]+2\sqrt{1-F^{m'}}}\\
\nonumber
&\ge&\cfrac{1}{2}+\delta-\cfrac{(3+2\delta)\sqrt{1-F^{m'}}}{2^{-k}+2\sqrt{1-F^{m'}}},
\end{eqnarray}
where we have used ${\rm Pr}[p=1]\ge2^{-k}$ to derive the last inequality.

\begin{figure}[t]
\includegraphics[width=7cm, clip]{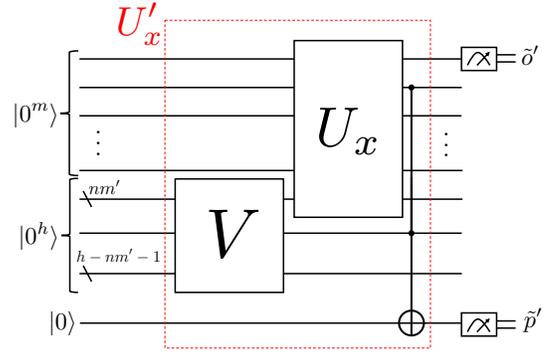}
\caption{By using the quantum circuit $V$ in Fig.~\ref{postgsfig}, we construct $U'_x$. By using this quantum circuit, we can solve any ${\sf postQMA}$ problem in quantum polynomial time with postselection, i.e., ${\sf postQMA}\subseteq{\sf postBQP}$. The output and postselection registers are denoted by $\tilde{o}'$ and $\tilde{p}'$, respectively.}
\label{postuxfig}
\end{figure}

On the other hand, when $x\notin L$, from ${\rm Pr}[o=1\ |\ p=1]\le1/2-\delta$,
\begin{eqnarray}
\nonumber
{\rm Pr}[o'=1\ |\ p'=1]&=&\cfrac{{\rm Pr}[o'=p'=1]}{{\rm Pr}[p'=1]}\\
\nonumber
&\le&\cfrac{{\rm Pr}[o=p=1]+2\sqrt{1-F^{m'}}}{{\rm Pr}[p=1]-2\sqrt{1-F^{m'}}}\\
\nonumber
&\le&\cfrac{(1/2-\delta){\rm Pr}[p=1]+2\sqrt{1-F^{m'}}}{{\rm Pr}[p=1]-2\sqrt{1-F^{m'}}}\\
\nonumber
&=&\cfrac{1}{2}-\delta+\cfrac{(3-2\delta)\sqrt{1-F^{m'}}}{{\rm Pr}[p=1]-2\sqrt{1-F^{m'}}}\\
\nonumber
&\le&\cfrac{1}{2}-\delta+\cfrac{(3-2\delta)\sqrt{1-F^{m'}}}{2^{-k}-2\sqrt{1-F^{m'}}}.
\end{eqnarray}}
Since $1-F^{m'}=\Theta(2^{-4k})$, $(3+2\delta)\sqrt{1-F^{m'}}/(2^{-k}+2\sqrt{1-F^{m'}})=O(2^{-k})$ and $(3-2\delta)\sqrt{1-F^{m'}}/(2^{-k}-2\sqrt{1-F^{m'}})=O(2^{-k})$.

The remaining task is to show that the success probability ${\rm Pr}[\tilde{p}'= 1]$ of postselection of $U'_x$ is at least the inverse of an exponential, which is required in the definition of ${\sf postBQP}$.
Since ${\rm Pr}[p'=1]\ge {\rm Pr}[p=1]-2\sqrt{1-F^{m'}}$ holds, ${\rm Pr}[\tilde{p}'= 1]={\rm Pr}[p''=1]{\rm Pr}[p'=1]\ge r^{m'}({\rm Pr}[p=1]-2\sqrt{1-F^{m'}})=\Omega(2^{-k}r^{m'})$.
As a result, we can conclude that if the quantum circuit $W$ exists, then ${\sf postQMA}\subseteq{\sf postBQP}$.
\hspace{\fill}$\blacksquare$

${\sf PP}={\sf PSPACE}$ leads to the first-level collapse of the counting hierarchy, i.e., ${\sf CH}={\sf PP}$.
This is because from ${\sf CH}\subseteq{\sf PSPACE}$,
\begin{eqnarray*}
{\sf PP}\subseteq{\sf CH}\subseteq{\sf PSPACE}={\sf PP}.
\end{eqnarray*}
Since ${\sf CH}={\sf PP}$ is unlikely {\red as discussed in Sec.~\ref{II}}, Theorem~\ref{additive} is evidence supporting the conclusion that generation of ground states is impossible even for postselected universal quantum computers.

Theorem~\ref{additive} is interesting, because it means that although ${\sf QMA}\subseteq{\sf postBQP}$~\cite{MW05}, generating ground states of {\red a given} $3$-local {\red Hamiltonian} seems to be beyond the capability of ${\sf postBQP}$ machines {\red in the worst case}.
In other words, generating ground states of any $3$-local Hamiltonians is just a sufficient condition to solve ${\sf QMA}$ problems, but it should not be a necessary condition.

\medskip
{\red\section{Hardness of approximately generating probability distributions}
\label{IV}}
Here, we will focus on the output probability distribution $\{p_z\}_z$ in Fig.~\ref{multiplicativefig} (a).
The proof of Theorem~\ref{additive} implies that given the values of $m$ and $m'$, and the classical descriptions of $H_x$ and $U_x$, it is hard to approximate $\{p_z\}_z$ with an exponentially small additive error $c'$ by using postselected quantum computation.
Therefore, the hardness with multiplicative error $1+c'$ also holds.
Here, we say that a probability distribution $\{p_z\}_z$ is generated with multiplicative error $c$ if and only if there exists a probability distribution $\{q_z\}_z$ such that $p_z/c\le q_z\le cp_z$ for any $z$.
When $p_z/(1+c')\le q_z\le (1+c')p_z$ holds for all $z$, the inequality $\sum_z|p_z-q_z|\le c'$ also holds.
Therefore, if we can show the hardness with additive error $c'$, then the hardness with multiplicative error $1+c'$ is also shown automatically.
In short, by using the argument used in the proof of Theorem~\ref{additive}, we can show the hardness with multiplicative error $1+c'$, which is exponentially close to $1$, for postselected quantum computation.
Hereafter, we will use a different argument to show the hardness with multiplicative error $1\le c<\sqrt{2}$, i.e., show that {\red in the worst case,} it is hard for postselected quantum computation to prepare approximate ground states from which we can generate $\{p_z\}_z$ in Fig.~\ref{multiplicativefig} (a) with multiplicative error $1\le c<\sqrt{2}$ given the success of the postselection.

\begin{figure}[t]
\includegraphics[width=9cm, clip]{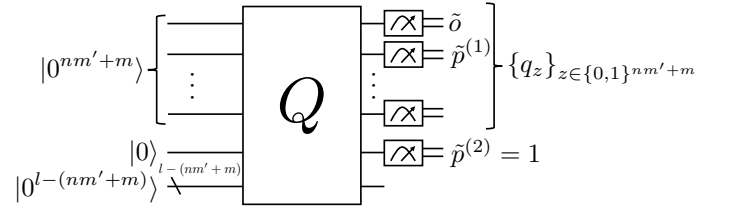}
\caption{A quantum circuit $Q$ generates the output probability distribution $\{p_z\}_{z\in\{0,1\}^{nm'+m}}$ with multiplicative error $c$ when the second postselection register $\tilde{p}^{(2)}=1$, which occurs with probability of at least the inverse of an exponential. In other words, $p_z/c\le q_z\le cp_z$ for any $z$. The symbols $\tilde{o}$ and $\tilde{p}^{(1)}$ are the output and first postselection registers of $Q$, respectively.}
\label{additivefig}
\end{figure}

The following theorem is our second main result:
\begin{theorem}
\label{inferiority}
Suppose that it is possible to, for any $n$-qubit 3-local Hamiltonian $H$, polynomials $m$ and $m'$, and $(nm'+m)$-qubit polynomial-size quantum circuit $U$, construct an $(l+1)$-qubit polynomial-size quantum circuit $Q$ for some polynomial $l(\ge nm'+m)$ in classical polynomial time, such that $Q$ takes $|0^{l+1}\rangle$ and generates the distribution $\{p_z\}_{z\in\{0,1\}^{nm'+m}}$ with multiplicative error $1\le c<\sqrt{2}$ when the postselection succeeds (i.e., $\tilde{p}^{(2)}=1$ in Fig.~\ref{additivefig}), where $p_z\equiv|\langle z|U(|0^m\rangle|g\rangle^{\otimes m'})|^2$ for any $z\in\{0,1\}^{nm'+m}$, $|g\rangle$ is a ground state of $H$, and ${\rm Pr}[\tilde{p}^{(2)}=1]\ge 2^{-k'}$ for a polynomial $k'$.
Then ${\sf PP}={\sf PSPACE}$.
\end{theorem}
{\red{\it Proof.}
We will use a similar technique as in Ref.~\cite{BJS10}. Let $L$ be in ${\sf postQMA}$.
From Lemma~\ref{remark}, for any instance $x$, there exist polynomials $m$ and $m'$ such that a polynomial-size quantum circuit $U_x$ with input $|0^m\rangle|g\rangle^{\otimes m'}$ efficiently decides whether $x\in L$ or $x\notin L$ under postselection of $p=1$ [see Fig.~\ref{multiplicativefig} (a)].
Here, $|g\rangle$ is a ground state of an $n$-qubit $3$-local Hamiltonian $H_x$ that depends on the instance $x$, and $n$ is a polynomial in $|x|$.
Let $p_z\equiv|\langle z|U_x(|0^m\rangle|g\rangle^{\otimes m'})|^2$ be the probability of the quantum circuit $U_x$ outputting $z$.
Let $o$ be the output register of $U_x$.
From the definition of ${\sf postQMA}$, when $x\in L$,
\begin{eqnarray*}
\cfrac{\sum_{z'\in\{0,1\}^{nm'+m-2}}p_{o=1,p=1,z'}}{\sum_{o\in\{0,1\},z'\in\{0,1\}^{nm'+m-2}}p_{o,p=1,z'}}\ge\cfrac{1}{2}+\delta
\end{eqnarray*}
for some constant $0<\delta<1/2$.
On the other hand, when $x\notin L$,
\begin{eqnarray*}
\cfrac{\sum_{z'\in\{0,1\}^{nm'+m-2}}p_{o=1,p=1,z'}}{\sum_{o\in\{0,1\},z'\in\{0,1\}^{nm'+m-2}}p_{o,p=1,z'}}\le\cfrac{1}{2}-\delta.
\end{eqnarray*}
Note that we can make the value of $\delta$ arbitrarily close to $1/2$ by increasing $m'$.

From the assumption with the Hamiltonian $H_x$, the quantum circuit $U_x$, and the polynomials $n$, $m$, and $m'$ described above, there exists a polynomial $l$ such that it is possible to efficiently construct an $(l+1)$-qubit quantum circuit $Q$ for the quantum circuit $U_x$ and the Hamiltonian $H_x$ (see Fig.~\ref{additivefig}).
By using the quantum circuit $Q$, the probability distribution $\{q_z\}_{z\in\{0,1\}^{nm'+m}}$ such that $p_z/c\le q_z\le cp_z$ for any $z$ can be efficiently generated when the second postselection register $\tilde{p}^{(2)}=1$.
From this inequality, the inequality $(\sum_{z\in S}p_z)/c\le \sum_{z\in S}q_z\le c\sum_{z\in S}p_z$ also holds for any subset $S$ of $\{0,1\}^{nm'+m}$.
Let $\tilde{o}$ and $\tilde{p}^{(1)}$ be the output and first postselection registers of $Q$, respectively.
Therefore, we obtain ${\rm Pr}[o=1\ |\ p=1]/c^2\le{\rm Pr}[\tilde{o}=1\ |\ \tilde{p}^{(1)}=\tilde{p}^{(2)}=1]\le c^2{\rm Pr}[o=1\ |\ p=1]$.
For any $c\in[1,\sqrt{2})$, we can find $\delta\in(0,1/2)$ such that $1\le c^2<1+2\delta$ by increasing $m'$.
When $x\in L$,
\begin{eqnarray*}
\label{yes}
{\rm Pr}[\tilde{o}=1\ |\ \tilde{p}^{(1)}=\tilde{p}^{(2)}=1]\ge\cfrac{1}{c^2}\left(\cfrac{1}{2}+\delta\right)>\cfrac{1}{2}.
\end{eqnarray*}
On the other hand, when $x\notin L$,
\begin{eqnarray*}
\label{no}
{\rm Pr}[\tilde{o}=1\ |\ \tilde{p}^{(1)}=\tilde{p}^{(2)}=1]\le c^2\left(\cfrac{1}{2}-\delta\right)<\cfrac{1}{2}-2\delta^2.
\end{eqnarray*}
From these two inequalities, we can conclude that if $Q$ exists for $1\le c<\sqrt{2}$ and any instance $x$, then ${\sf postQMA}\subseteq{\sf postBQP}$.
Note that postselecting two registers is allowed in postselected quantum computation because it can be reduced to one by using a single ancillary qubit $|0\rangle$ and the Toffoli gate, as in Fig.~\ref{postuxfig}.
Therefore, ${\sf PP}={\sf PSPACE}$.
\hspace{\fill}$\blacksquare$

In this proof, we considered the case where all $nm'+m$ qubits in Fig.~\ref{multiplicativefig} (a) are measured.
However, the same argument holds even when the number of measured qubits is less than $nm'+m$ as long as $o$ and $p$ are measured.}

\medskip
{\red\section{Conclusion and discussion}
\label{V}}
We have shown that {\red in the worst case,} efficient generation of ground states of {\red a given} $3$-local {\red Hamiltonian} is impossible for postselected quantum computation under a plausible assumption, i.e., the infiniteness of ${\sf CH}$.
So far, the quantum-computational-supremacy approach has been used only to show a quantum advantage.
Our results show that a similar approach can be used to show the opposite result, i.e., a quantum {\red limitation}.

Our argument essentially relies on the exponentially small {\red promise} gap between ground-state {\red energies} of two (families of) Hamiltonians each of which corresponds to YES and NO instances of a ${\sf PSPACE}$-complete problem{\red.
This gap is different from the spectral gap that is a gap between the ground-state and the first-excited-state energy of a Hamiltonian.
The spectral gap is related to the hardness of efficiently certifying ground states, while the promise gap is related to that of efficiently preparing ground states.
The certification is a task to check whether a given quantum state is close to the ideal ground state.
In Refs.~\cite{HKSE17,TM18}, concrete certification protocols have been proposed, and their efficiency depends on the spectral gap.
Especially when the spectral gap is exponentially small, they take exponential time (except for unnatural special cases).}

{\red Our argument does not work when the promise gap is polynomially small.
We leave, as a main open problem, whether efficiently preparing ground states of such Hamiltonians is still hard for postselected quantum computation.
Indeed, ${\sf QMA}\subseteq{\sf PP}$ has been shown in Ref.~\cite{MW05} without constructing witnesses of ${\sf QMA}$ by using a postselected quantum computer.
In other words, ${\sf QMA}\subseteq{\sf PP}$ does not mean that it is possible for postselected quantum computation to efficiently generate ground states of local Hamiltonians with the polynomially small promise gap.}

{\red On the other hand, our} argument can also be used to give evidence of the existence of at least one $3$-local Hamiltonian whose ground state cannot be {\red specified by} using a polynomial number of bits.
We say that a ground state is {\red specified by} using a polynomial number of bits if there exists a polynomial-size quantum circuit that outputs the ground state given the success of the postselection{\red.
By regarding these bits as a classical witness of ${\sf postQCMA}$, if there exists no such Hamiltonian, we can obtain ${\sf postQCMA}={\sf postQMA}$ and thus give the evidence.
Therefore, we would like to specify the ground state in such a way that the verifier in ${\sf postQCMA}$ can efficiently obtain the ground state from its specification.
When the Hamiltonian $H$ has a unique ground state, the ground state can also be specified by merely specifying a classical description of $H$.
However, the verifier in ${\sf postQCMA}$ cannot efficiently obtain the ground state from the classical description of $H$.}
Since ${\sf postQCMA}={\sf NP}^{\sf PP}$~\cite{MN17} and ${\sf postQMA}={\sf PSPACE}$~\cite{MN17}, if any ground state can be {\red specified by} using a polynomial number of bits, then ${\sf PSPACE}={\sf NP}^{\sf PP}$.
Therefore, from ${\sf CH}\subseteq{\sf PSPACE}$, the relation ${\sf PP}^{\sf PP}\subseteq{\sf CH}\subseteq{\sf PSPACE}={\sf NP}^{\sf PP}\subseteq{\sf PP}^{\sf PP}$ holds.
This means that the counting hierarchy collapses to its second level, i.e., ${\sf CH}={\sf PP}^{\sf PP}$.
{\red As an important point, we do not require the uniformity for the quantum circuit in this argument, while it is required in Theorems~\ref{additive} and \ref{inferiority}.
In Theorem~\ref{additive}, we suppose that there exists a quantum circuit $W$ generating ground states such that (i) its size is polynomial, and (ii) it can be constructed in classical polynomial time.
Here, we consider only condition (i).
In short, the quantum circuit considered here is of polynomial size, but it may be hard to find how to construct it for classical computers (more precisely, deterministic Turing machines).}

As an outlook, it would be interesting to strengthen the unlikeliness obtained from the efficient generation of ground states.
One direction is to improve the first-level collapse of ${\sf CH}$ to the zeroth-level one, i.e., ${\sf CH}={\sf P}$, which implies ${\sf P}={\sf NP}$.
Furthermore, it would also be interesting to reduce the number of measurements required to show Theorem~\ref{inferiority}.
Our argument needs at least two measurements ($o$ and $p$){\red.}
Can we reduce it to one?
Regarding Theorem~\ref{additive}, it would be interesting to consider whether we can show hardness for a constant fidelity.
As a common outlook among our results, it is open whether our results can be generalized to $2$-local Hamiltonians{\red.
This is because it is unknown whether the precise $2$-local Hamiltonian problem is ${\sf postQMA}$-complete~\cite{FL182}.
Here, the precise $2$-local Hamiltonian problem is the one of deciding whether the ground-state energy of a given $2$-local Hamiltonian is low or high with exponential accuracy (for the formal definition, see Ref.~\cite{FL18}).}

\medskip
\section*{ACKNOWLEDGMENTS}
We thank Tomoyuki Morimae for fruitful discussions.
{\red This work is supported by JST [Moonshot R\&D -- MILLENNIA Program] Grant Number JPMJMS2061.}
Y. Takeuchi is supported by MEXT Quantum Leap Flagship Program (MEXT Q-LEAP) Grant Number JPMXS0118067394 {\red and JPMXS0120319794}.
{\red ST is supported by the Grant-in-Aid for Transformative Research Areas No.JP20H05966 of JSPS.}

\medskip
\section*{\red APPENDIX: PROOF OF LEMMA~\ref{remark}}
{\red We give a proof of Lemma~\ref{remark}.

{\it Proof.}
Let $L$ be a language in ${\sf postQMA}$.
We show that for any instance $x$, there exists $|g\rangle^{\otimes m'}$ such that it is possible to decide whether $x\in L$ or $x\notin L$ in quantum polynomial time with postselection if $|g\rangle^{\otimes m'}$ is given.
Here, $|g\rangle$ is a ground state of a $3$-local Hamiltonian whose classical description can be generated in polynomial time from the instance $x$, and $m'$ is a polynomial in the length $|x|$.
To this end, it is sufficient to show that the YES-case witness $\rho_x$ in the definition of ${\sf postQMA}$ can always be replaced with $|g\rangle^{\otimes m'}$.

First, we define the complexity class ${\sf QMA}(c,s)$ as follows:
\begin{definition}
A language $M$ is in ${\sf QMA}(c,s)$ if and only if there exist polynomials $n$, $m$, and $k$, and a uniform family $\{U_y\}_y$ of polynomial-size quantum circuits, where $y$ is an instance, and $U_y$ takes an $n$-qubit state $\rho$ and ancillary qubits $|0^m\rangle$ as inputs, such that (i) if $y\in M$, then there exists a witness $\rho_y$ such that ${\rm Pr}[o=1\ |\ \rho_y]\ge c$ with a single-qubit output register $o$, and (ii) if $y\notin M$, then for any $\rho$, ${\rm Pr}[o=1\ |\ \rho]\le s$.
Here, ``polynomials" mean the ones in the length $|y|$ of the instance $y$.
\end{definition}
From ${\sf postQMA}={\sf PSPACE}$~\cite{MN17} and ${\sf PSPACE}=\bigcup_{r'\in{\rm poly}(|y|)}{\sf QMA}(1/2+2^{-r'},1/2-2^{-r'})$~\cite{FL18}, where ${\rm poly}(|y|)$ is a set of polynomials in $|y|$, any ${\sf postQMA}$ problem can be efficiently reduced to a problem in ${\sf QMA}(1/2+2^{-r},1/2-2^{-r})$ with a certain polynomial $r$ that depends on the original ${\sf postQMA}$ problem and the reduction method.

Next, we show that there exists a certain polynomial $\tilde{r}$ such that each problem in ${\sf QMA}(1/2+2^{-r},1/2-2^{-r})$ is polynomial-time reducible to a problem in ${\sf QMA}(1/2+2^{-\tilde{r}},1/2-2^{-\tilde{r}})$ whose YES-case witness $\rho_y$ is a ground state of a $3$-local Hamiltonian $H_y$ that depends on the instance $y$.
In other words, by reducing a problem in ${\sf QMA}(1/2+2^{-r},1/2-2^{-r})$ to another problem in ${\sf QMA}(1/2+2^{-\tilde{r}},1/2-2^{-\tilde{r}})$, we can always assume that the YES-case witness is a ground state.
To this end, we consider a ${\sf PSPACE}$-complete problem, the so-called precise $3$-local Hamiltonian problem~\cite{FL18}, which is a problem deciding whether the ground-state energy of a given $3$-local Hamiltonian $H_y$ is at most $a$ or at least $b$ under the condition that $b-a$ is at least the inverse of an exponential.
Any problem in ${\sf QMA}(1/2+2^{-r},1/2-2^{-r})$ can be efficiently reduced to the precise $3$-local Hamiltonian problem with $a=0$.
In other words, when $y\in M$ $(y\notin M)$, the ground-state energy of $H_y$ is at most $0$ (at least $b$).

However, since the ground-state energy of $H_y$ may be negative when $y\in M$, this reduction may be inconvenient for our purpose.
Remember that the operator norm $||H_y||$ is upper-bounded by $t$ (see Lemma~1).
Let $E_{\rm min}$ be the ground-state energy of $H_y$.
When $y\in M$, the inequality $-t\le E_{\rm min}\le 0$ holds.
On the other hand, when $y\notin M$, the inequality $b\le E_{\rm min}\le t$ holds.
To make the ground-state energy non-negative even when $y\in M$, we consider a scaled Hamiltonian $H'_y\equiv(H_y+tI^{\otimes n})/2$, where $I$ is the two-dimensional identity operator.
When $y\in M$ $(y\notin M)$, the ground-state energy of $H'_y$ is at most $t/2$ (at least $t/2+b/2$).
Note that a ground state of $H'_y$ is the same as that of $H_y$.

From Refs.~\cite{KSV02,AN02}, we can construct a polynomial-size quantum circuit $U_y$ that outputs $1$ with probabilities at least $1/2-b'/3$ and at most $1/2-2b'/3$ when the ground-state energies are at most $t/2$ and at least $t(1/2+b')$, respectively, if $|g\rangle$ is given, where $b'\equiv b/(2t)$.
Note that $U_y$ is an approximate quantum circuit of  that in Ref.~\cite{AN02} with exponential precision, because we restrict gate sets to approximately universal ones.
This restriction is necessary to use the equality ${\sf PP}={\sf postBQP}$.
By using $U_y$, we construct another polynomial-size quantum circuit $V_y$ such that it simulates $U_y$ with probability $1/(1+b')$, and otherwise always outputs $1$.
When $y\in M$, the quantum circuit $V_y$ outputs $1$ with probability at least $1/2+b'/[6(1+b')]$.
On the other hand, when $y\notin M$, the quantum circuit $V_y$ outputs $1$ with probability at most $1/2-b'/[6(1+b')]$.
Therefore, by setting $2^{-\tilde{r}}=b'/[6(1+b')]$, we conclude that there exists a polynomial $\tilde{r}$ such that each problem in ${\sf QMA}(1/2+2^{-r},1/2-2^{-r})$ is polynomial-time reducible to a problem in ${\sf QMA}(1/2+2^{-\tilde{r}},1/2-2^{-\tilde{r}})$ whose YES-case witness $\rho_y$ is a ground state of $H_y$.

Finally, in Ref.~\cite{MN17}, for any polynomial $r'$, it has been shown that any problem in ${\sf QMA}(1/2+2^{-r'},1/2-2^{-r'})$ can be solved in quantum polynomial time with postselection if polynomially many copies of a YES-case witness of the problem are given.
Therefore, $|g\rangle^{\otimes m'}$ can be used as a YES-case witness of any language in ${\sf postQMA}$.
\hspace{\fill}$\blacksquare$}

\end{document}